\documentclass[conference,9pt]{IEEEtran}
\IEEEoverridecommandlockouts
\usepackage{todonotes}
\usepackage{caption}
\usepackage{cite}
\usepackage{graphicx}
\usepackage{textcomp}
\usepackage{multirow}
\usepackage{xcolor}
\usepackage{tcolorbox}
\tcbuselibrary{listings,breakable}
\usepackage{listings}
\usepackage{color,soul}
\definecolor{codegreen}{rgb}{0,0.6,0}
\definecolor{codegray}{rgb}{0.5,0.5,0.5}
\definecolor{codepurple}{rgb}{0.58,0,0.82}
\definecolor{backcolour}{rgb}{1,1,1}

\usepackage{siunitx}

\usepackage{newfloat}

\DeclareFloatingEnvironment[
    fileext=lop,
    listname={List of Prompts},
    name=Prompt
]{prompt}

\DeclareFloatingEnvironment[
    fileext=lor,
    listname={List of Replies},
    name=Reply
]{reply}
\usepackage{xifthen}
\usepackage{circledsteps}



\newtcolorbox{userbox}{
  colback=cyan!5!white,
  colframe=cyan!60!black,
  boxrule=1pt,
  title=User Prompt,
}

\newtcolorbox{chatgptbox}{
  colback=green!5!white,
  colframe=green!50!black,
  boxrule=1pt,
  title=AI Reply,
}

\newenvironment{llmprompt}[2]
{%
  \begin{prompt}[t]
  \footnotesize
\tcbset{
  breakable=false,
  boxsep=1pt,
  top=1pt,
  bottom=1pt
}
\setlength{\abovecaptionskip}{2pt}
  \def\llmcaption{#2}
  \def\llmlabel{#1}
  \begin{userbox}
}
{%
  \end{userbox}
  \caption{\llmcaption} \label{llm:\llmlabel}
  \end{prompt}
}

\newenvironment{llmreply}[2]
{%
  \begin{reply}[t]
  \footnotesize
\tcbset{
  breakable=false,
  boxsep=1pt,
  top=1pt,
  bottom=1pt
}
\setlength{\abovecaptionskip}{2pt}
  \def\llmcaption{#2}
  \def\llmlabel{#1}
  \begin{chatgptbox}
}
{%
  \end{chatgptbox}
  \caption{\llmcaption} \label{llm:\llmlabel}
  \end{reply}
}

\usepackage{orcidlink}
\usepackage{listings}
\usepackage{caption}
\usepackage{hyperref}
\usepackage{lipsum}
\usepackage{multirow}
\usepackage{tikz}
\hypersetup{
    colorlinks,
    citecolor=black,
    filecolor=black,
    linkcolor=black,
    urlcolor=black
}
\definecolor{codegreen}{rgb}{0,0.6,0}
\definecolor{codegray}{rgb}{0.5,0.5,0.5}
\definecolor{codepurple}{rgb}{0.58,0,0.82}
\definecolor{backcolour}{rgb}{1,1,1}
\lstdefinestyle{mystyle}{
    language=Python,
    backgroundcolor=\color{backcolour},   
    commentstyle=\color{blue},
    keywordstyle=\color{codepurple},
    numberstyle=\scriptsize\color{gray},
    framexleftmargin=1.5em,
    xleftmargin=1.5em,
    stringstyle=\color{codegreen},
    basicstyle=\ttfamily\tiny,
    breakatwhitespace=false,         
    breaklines=true,                 
    captionpos=b,                    
    keepspaces=true,                 
    numbers=left,
    numbersep=3pt,
    tabsize=2,
    showspaces=false,                
    showstringspaces=false,
    showtabs=false,                  
    tabsize=1
}
\usepackage{amsmath,amsfonts}
\usepackage{algorithmic}
\usepackage{graphicx}
\usepackage{textcomp}
\usepackage{xcolor}
\usepackage{hyperref}
\usepackage{cleveref}
\usepackage{enumitem}
\usepackage{subcaption}
\Crefformat{figure}{#2Fig.~#1#3}
\usepackage[nolist]{acronym}

\def\BibTeX{{\rm B\kern-.05em{\sc i\kern-.025em b}\kern-.08em
    T\kern-.1667em\lower.7ex\hbox{E}\kern-.125emX}}
\title{Automated Multi-Line Bug Repair Using Model Checker Aided LLMs:  A Case Study on RISC-V Processors} 

\author{Debleena Das\IEEEauthorrefmark{1}\IEEEauthorrefmark{4}, Chandan Kumar Jha\IEEEauthorrefmark{2}, Muhammad Hassan\IEEEauthorrefmark{2}\IEEEauthorrefmark{3}, Rolf Drechsler\IEEEauthorrefmark{2}\IEEEauthorrefmark{3},\\ Ansuman Banerjee\IEEEauthorrefmark{1}, Subhra K. Das\IEEEauthorrefmark{5}, Amlan Chakrabarti\IEEEauthorrefmark{4} \\\IEEEauthorblockA{
  \IEEEauthorrefmark{1} Indian Statistical Institute, India,
  \IEEEauthorrefmark{2}University of Bremen, Germany,
  \IEEEauthorrefmark{3}DFKI GmbH, Germany\\   
 \IEEEauthorrefmark{4} A. K. Choudhury School of Information Technology, University Of Calcutta, India,
 \IEEEauthorrefmark{5} Thales India Private Limited, India\\
 }
}
\begin{document}
\bstctlcite{IEEEexample:BSTcontrol}
\maketitle
\begin{abstract}
Functional verification of \emph{Register-Transfer Level} (RTL) designs has become a critical bottleneck in modern semiconductor development, often consuming the majority of project resources.
Debugging, in particular, remains an especially time-consuming task, requiring careful analysis of simulation traces or counterexamples.
Recent work has explored applying \emph{Large Language Models} (LLMs) to RTL debug and repair with great promise.
However, the existing approaches typically are for single-line bugs, operate in open-loop, rely heavily on prompts or retrieval, lack scalability, and often produce incorrect fixes.
In this paper, we propose a novel automated RTL bug repair methodology that combines LLMs, assertions, and a model checker to provide scalable, trustworthy semantic grounding and formal guarantees.
Concretely, the methodology leverages failing assertions to guide the LLM in generating patches for multi-line bugs.
Each candidate patch is formally verified using a model checker (JasperGold), ensuring only valid repairs are accepted.
We evaluate the efficacy and scalability of our methodology on the BRISC-V RISC-V core using mutation testing, introducing several mutation types,
and demonstrate that our methodology can repair up to $98\%$ of single-line RTL bugs and 93\% of the multi-line bugs.
To the best of our knowledge, this is the first work to show automated, LLM-assisted RTL multi-line bug repair results on a complex RISC-V processor design in a scalable manner.
\end{abstract}

\maketitle

\section{Introduction}

\noindent
Functional verification of \emph{Register-Transfer Level} (RTL) designs has emerged as a principal bottleneck in the semiconductor development process, often consuming up to 70\% of overall project resources~\cite{foster2015trends,SHI:2022}.
This growing burden is largely driven by the increasing complexity and functional richness of modern digital systems, which incorporate deep pipelines, multiple interacting sub-systems, and a variety of protocols and timing constraints.
As designs scale in size and sophistication, the verification effort required to ensure correctness and safety has grown disproportionately.
Among the various verification tasks, debugging, i.e., identifying and correcting functional errors in \emph{Hardware Description Language} (HDL) code remains particularly time-consuming.
It often requires manual analysis of simulation traces or counterexamples, and demands a detailed understanding of both the design intent and implementation~\cite{klemmer2024waving,nadeem2025polynomial}.

In recent times, \emph{Large Language Models} (LLMs) have shown success in software engineering tasks such as code generation and automated program repair.
This success has motivated research on applying LLMs to hardware design tasks~\cite{hdldesign,llm4dv}, with studies exploring RTL code generation~\cite{makecount,verilogcoder,verigen}, security bug fixes~\cite{hardsecurity}, and syntax error correction~\cite{rtlfixer,LLMvtd}. 

In particular, a number of recent research articles have reported efforts on using LLMs to generate \emph{SystemVerilog Assertions} (SVAs) from various inputs such as natural-language specifications, source code comments, or even the RTL code itself~\cite{vert,kande,laagrvllm,rtltosva, sangam}.
Frameworks like \emph{AssertLLM} and \emph{Spec2Assertion} translate design specifications into formal properties, easing the burden of writing assertions manually~\cite{assertllm,spec2assert}.
Other research has demonstrated iterative, human-in-the-loop flows where feedback from a formal verification tool is used to refine LLM-generated assertions~\cite{LISA}, using techniques like \emph{Retrieval Augmented Generation} (RAG), \emph{Chain-Of-Thought} (COT), and agentic frameworks~\cite{aspdac26,jha2025large}.
Although these advances streamline the creation of verification artifacts, they do not directly address the task of repairing a buggy RTL.

There have been recent attempts to repair RTL using LLM-based methods.

\emph{AssertSolver}~\cite{assertSolver} trains an LLM on a large synthetic dataset of bugs, failure logs, and fixes, enabling it to predict code patches from a given assertion failure log.
These methods can suggest plausible fixes, but rely on a static pre-trained model and perform the repair in a single pass.
~\textcolor{black}{The HDLdebugger framework introduces a retrieval-augmented fine-tuning approach, and along with HDL documents, the LLM takes the buggy code, its error message, task prompt, and code RAG, and the edit distance between the correct code and the buggy code is their evaluation metric~\cite{hdldebugger}.}
Other bug-patching strategies~\cite{qayyum2024bugs, ACMTODES-khushboo} employ an iterative ``bug closure'' process using 
RAG to incorporate design specifications.
If the initially retrieved context is insufficient, these methods repeatedly pull in more specification content or code context and re-prompt the LLM. This iterative method can eventually find a fix for up to 60\% of the reported bugs.

While these methods have shown a lot of promise by fixing many of the reported bugs, these works focus on single-line bugs, lack trustworthy semantic grounding in a scalable manner, and typically run open-loop. Hence, the identification and fixes often tend to be plausible but wrong.

\begin{figure}
\centering
\includegraphics[width=\linewidth]{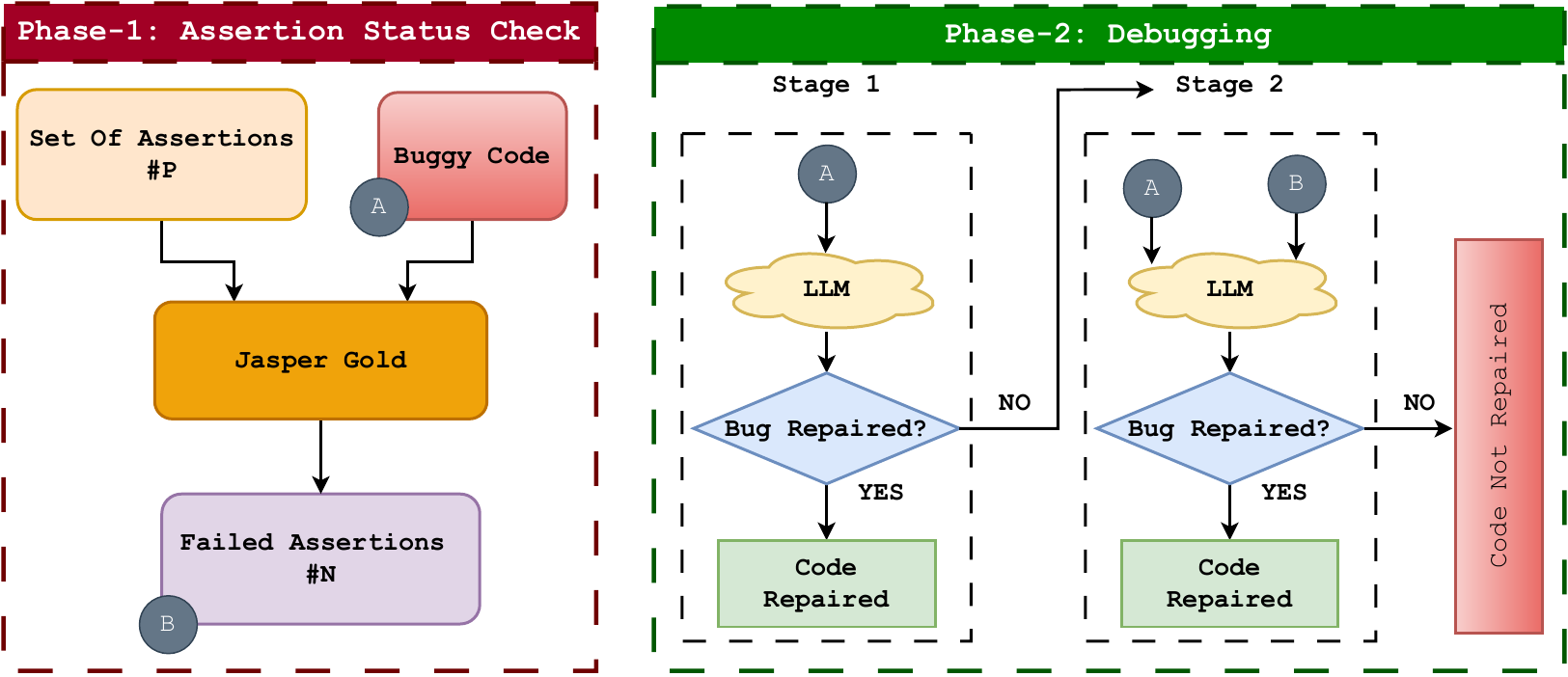}

\caption{\label{fig:systemarch} Workflow of LLM-assisted automated bug repair methodology}
\end{figure}

\textbf{Contribution:} We propose a novel automated RTL bug repair methodology for both single and multi-line bugs that uses LLMs, assertions, and a model checker to synthesise bug patches on a complex RISC-V processor in a scalable manner.
More concretely, the goal is to leverage the semantic structure of assertions and the output of a model checker to build a context for the LLM, so that it can propose an accurate fix.
Consequently, it provides a trustworthy grounding and formal guarantees as the model checker provides an objective oracle.

To this end, our methodology is two-fold:

First, the user asks the LLM to generate the patch only from the buggy code.
Second, we provide the failing assertions in addition to the buggy code to aid in the correction process.
In both steps, the proposed patch is always formally verified using a model checker.
For scalability, we conducted an extensive set of experiments on BRISC-V~\cite{BRISCV} where we took advantage of mutation testing~\cite{serrestou2007functional}.
Mutation testing allows for a comprehensive assessment of the quality and thoroughness of the verification
process by introducing controlled modifications in the given design.
We study $10$ mutation classes. $9$ mutations from the existing literature and $1$ new mutation, which injects an\emph{``if-else'' indentation bug}~\cite{ACMTODES-khushboo}. 
We inject the mutations in the RTL using LLM.
Our proposed method was able to fix upto $98\%$ of the single-line bugs and $93\%$ of the multi-line bugs certified by the model checker.
To the best of our knowledge, this is the first work to present an automated LLM-assisted repair result on a complex RISC-V processor code. 
Our contributions are summarized below:

\begin{itemize}[leftmargin=*]
\item We propose a novel LLM-assisted automated RTL bug repair approach with assertions and a model checker in the loop, which provides trustworthy semantic grounding and formal guarantees.
\item The methodology successfully fixes single-line and multi-line bugs, where the bug could be anywhere in the procedural blocks. 

\item We experiment using mutation testing with a set of $10$ mutation classes on an open source RISC-V IP, on different modules with a mix of auto-generated and hand-written assertions.

\item We implement an immediate prior art~\cite{ACMTODES-khushboo} (referred to as 
\textit{Stage~1} in our experiments) on LLM-aided 
RTL repair, and provide comparative results. As seen from our experiments, the presence of 
assertions indeed helps the fixing task.

\end{itemize}

\section{Motivating Example}
\label{sec:motivation}
In this section, we present a small example with a bug in RTL to highlight the problem in using open-loop methods, which do not have trustworthy grounding.
The example is a code snippet from the \emph{Cache Coherence Controller} (CCC) module inside BRISC-V.
The CCC typically implements one of the cache coherence protocols to ensure coherence and consistency of data shared across multiple cores~\cite{Hennessypatterson}. In the BRISC-V core used in this work, the MESI protocol is implemented.
In particular, the CCC code describes a finite state machine highlighting the different states of the snoop operation, with $7$ different states and respective transitions. 
This module has an assertion suite consisting of $56$ concurrent assertions, all of which evaluate to \emph{true} 
 when run using the model checker 
JasperGold~\cite{jasper}. 
We inject a conditional bug by inverting the \texttt{"if conditional"} statement. 
The original code was \texttt{"if (hit) begin"} and after the bug injection, it was changed to 
\texttt{"if (!hit) begin"}. 

As shown in Prompt~1, a verbose prompt was provided to the LLM to suggest a fix for the mutation.
\begin{llmprompt}{prompt2}{Role setup for the LLM in \textit{Stage~1}}
\label{prompt:user1}
    \textbf{Role:}
    You are an expert RTL assertion verification engineer with 25 years of experience in SystemVerilog HDL design. If you are given a hardware design in SystemVerilog, you can detect the bug and patch it.  
    
\textbf{Your task: 1}
Please identify the bug in the given block of code: ...
\end{llmprompt}
\begin{llmreply}{prompt1reply}{ LLM's reply to first prompt}
\label{prompt:Reply1}
 \begin{lstlisting}[aboveskip=0pt, belowskip=0pt,language=verilog]    
   Possible dead/looping path in WAIT_FOR_RESP when line_counter == ratio and intf_msg == MEM_RESP.
   
   (intf_msg == MEM_RESP) begin
  r_snoop_msg <= EN_ACCESS;
  r_snoop_address <= r_bus_address;
  ... clear r_snoop_data ...
  state <= WAIT_FOR_RESP;
end \end{lstlisting}     
 \end{llmreply}
In response to Prompt 1, the LLM was unable to identify the correct patch and produced a hallucination, as shown in Reply 1. This happened despite the simple nature of the single-line bug.
This highlights the limitation of only using the LLM for bug fixing. Prior works have shown that some of the bugs can be corrected by iteratively prompting the LLM. However, the approach is still non-deterministic and can require an arbitrary number of iterations. In addition, these analysis were limited to single-line bugs. In our work, we alleviate this limitation by using high-quality assertions and a model checker in the loop.

Prior works have relied solely on LLMs for the bug-fixing methodology, greatly limiting their applicability and potential. In this work, rather than only relying on LLMs, we show that having assertions and a model checker in the loop can largely aid and improve the bug-fixing process. We show that using the proposed methodology, we can fix bugs in larger circuits like a RISC-V processor, which has also been chosen as the case study in this work. In addition, while the prior works have mostly been limited to single-line bugs, we show that the proposed methodology is effective for both single-line and multi-line bugs.

\section{LLM-assisted Automated Bug Repair} 
\label{sec:proposed_methodology}
This section presents our LLM-assisted automated RTL bug repair methodology with assertions and a model checker in the loop, which provides trustworthy grounding and formal guarantees.
Fig.~\ref {fig:systemarch} shows a high-level workflow of our proposed 
methodology, which takes in RTL code and a set of assertions and carries out $2$ steps. First, we take the 
assertions and pass them to JasperGold, which runs the assertions on the code and produces the set of Failed 
Assertions. The second phase consists of $2$ stages, \textit{Stage~1}: the LLM is asked to repair the buggy code without any additional context. If the bug is not repaired, then in  \textit{Stage~2}: the set of failed assertions is provided to the LLM in addition to the code, and it is asked to repair the code. 
If the bug is repaired, then "Code Repaired" is reported, and if the bug is not repaired after \textit{Stage~2}, "Code not repaired" is reported. We now elaborate on each of the two \textit{Stages} in detail.

\subsection{Proposed Methodology}
Given an RTL design ${\mathcal R}$, a set of assertions ${\mathcal P}$, and a set of 
bug types, our objective is to be able to localize the bugs of each of the given types and generate  fixes with the help of an LLM and a formal verification tool.
We assume the bugs 
can be in any single-line or across multiple-lines anywhere in the RTL code. Also, all assertions in 
the set ${\mathcal P}$ hold on ${\mathcal R}$, when evaluated using the formal 
verifier, which evaluates an assertion on all possible scenarios before 
concluding a pass. 
Fig.~\ref {fig:systemarch} shows a high level workflow of our proposed 
methodology, comprising of $2$ main \textit{Phases}: 1) \textit{Phase-1:} Assertion Status Check, and 2) \textit{Phase-2:} Debugging.

\subsubsection{Phase-1 Assertion Status Check}
\noindent
\noindent
After obtaining the mutated piece of code from the LLM, it is inserted into 
${\mathcal R}$, on which the next steps are carried out. The mutated code is then run in JasperGold against ${\mathcal P}$. 
Due to the presence of single- or multiple-line mutations that act as bugs in this work, some assertions may result in \emph{counterexamples} (CEX), showing that these assertions were violated in the design. %

\subsubsection{Phase-2: Debugging}
\noindent
Our repair phase comprises $2$ \textit{Stages}, as discussed below.
After each \textit{Stage}, the patches suggested by the LLMs are examined by the JasperGold tool. If all assertions pass, we conclude that the bugs have been repaired. 

\begin{itemize}[leftmargin=*]
    \item {\em Stage 1:} We first provide the mutated code, i.e., the buggy code, and ask the LLM to identify the bugs. The LLM is only provided with the information that the code has bugs, and we want to repair the bug. No additional information is provided at this Stage to the LLM. Each patch generated is examined through the formal verifier against the mutated code, and the number of Failed assertions are counted. If the number of Failed assertions is reduced, then it is considered a correct patch suggested by the LLM. This can happen as in multi-line bugs, as some of the bugs can be corrected, while for the others we may need to go to the \textit{Stage 2}. 
    We proceed to the next step if the verifier reports that some assertions produce a violation leading to counterexamples. 

    \item {\em Stage 2:} Assume out of the set ${\mathcal P}$, a total of ${\mathcal N}$ assertions fail in the presence of single-line or multi-line mutations, i.e., bugs. 
    
    We add these failing assertions to the LLM in the same context window, asking for a patch as shown in Prompt~2. 
    Evidently, this brings in more context to the LLM to understand the 
    bug, in the context of the code block and the failing assertion(s). Note that in the prior state-of-the-art~\cite{ACMTODES-khushboo}, only the line with the bug was given to the LLM for correction. This further highlights the limitation of the prior approach, that the line with the bug needed to be identified before the LLM suggested a patch. However, in our work, we give the entire code block with the bug. Some patches are suggested by the LLMs that are checked to see, if all the assertions are now passing (as shown in Reply~2). If all the assertions are passing at this stage we say that the bugs in the design has been repaired.  
\end{itemize}

\noindent
Hence, we clearly see that using the proposed method, we not only are able to perform debugging on multi-line bugs, but we are also using the model checker to guarantee that the generated patch is indeed correct. In the following section, we present our experience on the BRISC-V code base 
with the mutations.

\subsection{Case Study: RISC-V Processor}

In this section, we revisit the motivating example from Section~\ref{sec:motivation} and illustrate the effectiveness of the proposed LLM-assisted bug repair methodology.
As mentioned earlier, the LLM was not able to patch the bug.
When the mutation was inserted in the Snooper RTL code inside the CCC, several 
assertions produced counterexamples. When one such failing assertion was 
passed on to the LLM, the nature of the reply changed. 
With the structured content of an assertion, the LLM gained more context, and it could identify and repair the bug as shown in Reply~2.
In this stage, we invoked the LLM with the prompt shown in Prompt~2, embedding one failing 
assertion inside to aid in its introspection.

\begin{llmprompt}{prompt2}{Role setup for the LLM}
\label{prompt:user2}
\textbf{Your Task: 2}
 Your provided solution is incorrect. Please identify the bugs in the following SystemVerilog code based on the given failed assertions on the buggy code. 
The list of failed assertions on the buggy code:
\begin{lstlisting}[aboveskip=0pt, belowskip=0pt,language=verilog]
  property rfobcast_hit_ratio;
  @(posedge clock) disable iff (reset)
    state == ACTION && r_bus_msg == RFO_BCAST && hit && (line_counter == ratio) |->
      ##1 state == WAIT_FOR_RESP && r_snoop_msg == EN_ACCESS && r_invalidate == 1;
endproperty
...
\end{lstlisting}
\end{llmprompt}
\begin{llmreply}{prompt1reply}{ LLM reply to second prompt}
\label{prompt:Reply2}
\textcolor{blue}{\textbf{Identified Bug}}
 \begin{lstlisting}[aboveskip=0pt, belowskip=0pt,language=verilog]
 In the ACTION RFO\_BCAST branch the hit test is inverted. The code uses if (!hit) where it should use if (hit).
 \end{lstlisting}    
 \textcolor{blue}{\textbf{Reason of the Bug}}
 \begin{lstlisting}
    When the cache has the line (hit) the FSM takes the non-hit path, so r_invalidate is not asserted.     
    That directly breaks the assertions that expect r_invalidate==1 and a transition to WAIT_FOR_RESP / INVALIDATE_LINE when hit is true.
\end{lstlisting}
\textcolor{blue}{\textbf{Patching}}
\begin{lstlisting}
RFO_BCAST: begin
  (hit) first
  if (hit) begin
     if (line_counter == ratio) begin 
     ...
\end{lstlisting}   
\end{llmreply}
The response received from the LLM is outlined in Reply~2. It clearly 
explains (a) the perceived cause of the bug, (b) the reason for the presence of the bug, and (c) a suitable patch to fix it. Indeed, the patch given in Reply~2 fixes the issue, thereby asserting the LLM's ability 
to clearly identify the context and the mutation. It was able to achieve this even though it was not explicitly informed about the location of the bug or the 
mutation type.
This reinforces the fact that the proposed methodology can patch the bugs correctly while providing formal guarantees at the same time. 
In addition, it advances the state of the art, where only the lines with bugs were given to the LLM for correction~\cite{ACMTODES-khushboo}. This case studied in the prior works is also impractical in certain scenarios, as they assume that the line with the bug is known a priori. Unlike prior work,  a section of code with the buggy line is given to the LLM in our methodology. We want to highlight that this simple example was chosen to highlight the efficacy of our methodology.

\section{Experimental Evaluation} \label{Sec:expts}
\noindent
In this section, we discuss the experimental setup, bug insertion, and the results obtained using the proposed methodology.
\subsection{Experimental Setup}
\noindent
The experimental evaluation is carried out on several modules of the BRISC-V core~\cite{briscv2}, an open source implementation of a parameterized, synthesizable, multicore architecture.
We use JasperGold~\cite{jasper} as the formal verification tool and GPT-{5.5}~\cite{gpt5.5} as the LLM.
All assertions used throughout the experiment, both before and after the mutation insertion, are verified using  JasperGold.

\subsubsection{Prompts}
For prompting the LLM, we use the OpenAI web-based ChatGPT interface with the GPT-5.5 model. This web-based version of the LLM instance can keep track of the code once it is assigned as a role, and we get the output iteratively, providing multiple tasks. The same can also be automated using the API calls. However, for this work, we have used the web interface. 
Initially, the code is given to the LLM, and then the role of the LLM is defined.
To set the context of the LLM, we set the roles of the LLM as an expert in RTL assertion verification engineering with 25 years of experience in SystemVerilog HDL design. If it is given a hardware design in SystemVerilog, it can detect the bug and patch it. 
With the role-playing setup, we try to improve responses aligned with assigned roles~\cite{tseng2024talespersonallmssurvey}.
In Section~\ref{sec:resultsandfind}, we discuss the results obtained for each of the \textit{Stages} in detail in Table~\ref{tab:results_main}. 


\subsubsection{Assertions}\label{sec:assertion} 
Assertions play a key role in our overall workflow. For some of the modules, 
 assertions were manually generated,
 while for others, assertions were 
generated using an LLM in a different session. 
The important metric of the generated assertion is the 
formal coverage of the assertion suite, which gives confidence in the 
proof span of the assertions with respect to the code statements, 
states, transitions, and faults. Given an RTL model and a set of assertions, 
JasperGold provides this information to a verification engineer, 
which guides one in assertion IP development. We utilize this feature 
crucially in our work. If the assertion coverage is low, it may not always 
be possible to get failing assertions that can aid in localization. 
Also, it may be noted that the generation of high coverage assertion IPs is a challenge. 
One of the important concerns in not being able to reproduce the experiments 
reported in~\cite{ACMTODES-khushboo} is the lack of useful assertions 
for the OpenTitan~\cite{opentitan} modules. 
Several LLM-assisted assertion generators report assertion generation 
capabilities on OpenTitan modules. In particular, we experimented with~\cite{sangam} and ~\cite{LISA}, but our experience was negative. Most of the assertions 
either failed or passed without any trend, making it impractical for us to reproduce their work. Hence, in this work, we used all the assertions that were manually generated and were evaluated to be of high quality using JasperGold. 

\subsubsection{BRISC-V modules used}

\begin{figure}[]
  \centering

  \begin{tcolorbox}[
    width=0.8\columnwidth,
    boxrule=0.65pt,
    colback=gray!5,
    colframe=blue!75!black,
    left=0pt,right=0pt,top=0pt,bottom=0pt,
    boxsep=1pt,
    title=Example of Nested If-else Mutation
  ]

  \footnotesize

  \begin{minipage}[t]{0.45\columnwidth}
    \begin{lstlisting}[title=Correct Code]
REQ_FLUSH:begin
    if(hit)begin
      if(dirty)begin
       ...
      end
      else begin 
      if(line_counter == ratio) begin
        ...
      end
      else begin
      ...
      end
    end
    \end{lstlisting}
  \end{minipage}
  \hspace{2mm}
  \begin{minipage}[t]{0.45\columnwidth}
    \begin{lstlisting}[title=Mutated Code]
REQ_FLUSH:begin
   if(hit)begin             
     ...
    end
    
   else if(dirty)begin
      if(line_counter == ratio)
      ....
      end
      else begin
      ...
      end
    end
    \end{lstlisting}
  \end{minipage}

  \end{tcolorbox}

  \caption{Correct and Mutated code before and after Nested If-else Mutation}
  \label{fig:figNestedIf}
\end{figure}
\begin{llmprompt}{prompt3}{Mutation Insertion Prompt}
\label{prompt:user3}
You are an expert in mutation generation of SystemVerilog code.
These are the types of mutations that you need to generate.

(Types of mutation described)

Task 1: Generate only one "Mutation type" Mutation in the original reference design and give me the output. 
\end{llmprompt}
\begin{table}[]
\centering
\caption{Various Classes of Mutations inserted in RTL code}
\resizebox{0.75\linewidth}{!}{
\begin{tabular}{|c|c|c|}
\hline
\textbf{S. No.} & \textbf{Mutation Type} & \textbf{Description} \\
    \hline

I  &  Logical & \begin{tabular}[c]{@{}l@{}} Change logical operations (e.g., $\&$ to $\|$)\end{tabular}\\
\hline
II & Relational & \begin{tabular}[c]{@{}l@{}}Change relational operators\\  (e.g., $==$ to $!=$)\end{tabular}\\
\hline
III & Arithmetic & Change arithmetic operators (e.g., $+$ to $-$)\\
\hline
IV & Constant & Change the constant value assigned to a variable\\
\hline
V & Conditional & Modify the 
conditions of the if-else statement\\
\hline
VI & Stuck at Fault & \begin{tabular}[c]{@{}l@{}}Assign a variable 
   from constant \\
   $0$ to $1$ or vice versa\end{tabular} \\
   \hline
   
VII & Sensitivity List & Change posedge to negedge and vice versa\\
   \hline
VIII & Connection & \begin{tabular}[c]{@{}l@{}} Change the interconnection  between\\   signals (swap) \end{tabular}  \\ 
 \hline
IX & Slicing & \begin{tabular}[c]{@{}l@{}} Change the number of bits used when \\ slicing is done \end{tabular}  \\ 
 \hline
X & {Nested if-else} & \begin{tabular}[c]{@{}l@{}}{ Change the Nested if-else alignment} \end{tabular}  \\ 
\hline
\end{tabular}
}
 \label{tab:mutable}
 \end{table}
\begin{table*}[]
\centering
\caption{Experiments on BRISC-V modules (FC: Formal Coverage, SC: Stimuli Coverage, CC: Checker Coverage, 1: Bug repaired in \textit{Stage 1}, 2: Bug repaired in \textit{Stage 2}, X: Bug not repaired after \textit{Stage 1 } and \textit{Stage 2}, $\--$: Bug type absent )}
\resizebox{0.75\linewidth}{!}{%
\begin{tabular}{|c|c|c|ccc|cccccccccc|}
\hline
Module & \# Lines & \#Assertions & \multicolumn{3}{c|}{Coverage} & \multicolumn{10}{c|}{Mutations}                           \\ \hline
&          &              & \multicolumn{1}{c|}{FC} & \multicolumn{1}{c|}{SC} & CC & \multicolumn{1}{c|}{I} & \multicolumn{1}{c|}{II} & \multicolumn{1}{c|}{III} & \multicolumn{1}{c|}{IV} & \multicolumn{1}{c|}{V} & \multicolumn{1}{c|}{VI} & \multicolumn{1}{c|}{VII} & \multicolumn{1}{c|}{VIII} & \multicolumn{1}{c|}{IX} & X \\ \hline

L1 bus interface &  $315$  &     $71$     & \multicolumn{1}{c|}{$89.06\%$}   & \multicolumn{1}{c|}{$95.31\%$}      & \multicolumn{1}{c|}{$93.75\%$}  & \multicolumn{1}{c|}{$1$}   & \multicolumn{1}{c|}{$1 $}    & \multicolumn{1}{c|}{$1$}   & \multicolumn{1}{c|}{$2 $}  & \multicolumn{1}{c|}{$1$}    &  \multicolumn{1}{c|}{$2 $}    &  \multicolumn{1}{c|}{$1$}    &  \multicolumn{1}{c|}{$2$}  & 
\multicolumn{1}{c|}{$2 $} & \multicolumn{1}{c|}{$2$}   \\ \hline

Coherence Controller &  $171$  &     $24$     & \multicolumn{1}{c|}{$91.00\%$}   & \multicolumn{1}{c|}{$99.00\%$}   &     \multicolumn{1}{c|}{$92.00\%$}  & \multicolumn{1}{c|}{$2 $}   
& \multicolumn{1}{c|}{$2$}    & \multicolumn{1}{c|}{$1$}   & \multicolumn{1}{c|}{$2$}  & \multicolumn{1}{c|}{$2$}   & \multicolumn{1}{c|}{$1 $}    &   \multicolumn{1}{c|}{$1$}    &  \multicolumn{1}{c|}{$2$}  & \multicolumn{1}{c|}{$\--$} &  \multicolumn{1}{c|}{$2$}   \\ \hline

Snooper &  $322$  &     $56$     & \multicolumn{1}{c|}{$90.00\%$}   & \multicolumn{1}{c|}{$90.63\%$}      & \multicolumn{1}{c|}{$98.75\%$}  & \multicolumn{1}{c|}{$2$}   & \multicolumn{1}{c|}{$1 $}    & \multicolumn{1}{c|}{$2$}   & \multicolumn{1}{c|}{$2 $}  & \multicolumn{1}{c|}{$1 $}   & \multicolumn{1}{c|}{$2 $}    &  \multicolumn{1}{c|}{$1$}    &  \multicolumn{1}{c|}{$2$}  &  \multicolumn{1}{c|}{$2$}  &  \multicolumn{1}{c|}{$2 $}  \\ \hline

Lxcache controller &  $276$  &   $76$       & \multicolumn{1}{c|}{$97.71\%$}   & \multicolumn{1}{c|}{$97.71\%$}   &    \multicolumn{1}{c|}{$100.00\%$}  & \multicolumn{1}{c|}{$1$}   & \multicolumn{1}{c|}{$2 $}    & \multicolumn{1}{c|}{$1$}   & \multicolumn{1}{c|}{$2$}  & \multicolumn{1}{c|}{$1$}   & \multicolumn{1}{c|}{$1$}    & \multicolumn{1}{c|}{$1 $}    &   \multicolumn{1}{c|}{$2 $}  & \multicolumn{1}{c|}{$2$} &  \multicolumn{1}{c|}{$2$}  \\ \hline

Lxbus interface &  $254$  &     $35$        & \multicolumn{1}{c|}{$92.73\%$}       & \multicolumn{1}{c|}{$93.64\%$}  & \multicolumn{1}{c|}{$99.09\%$}   & \multicolumn{1}{c|}{$\times$} 
& \multicolumn{1}{c|}{$1$}   & \multicolumn{1}{c|}{$1$}  & \multicolumn{1}{c|}{$2$}   & \multicolumn{1}{c|}{$2 $}    &  \multicolumn{1}{c|}{$1 $}    &  \multicolumn{1}{c|}{$1$}    &  \multicolumn{1}{c|}{$1$}  & \multicolumn{1}{c|}{$2$} & \multicolumn{1}{c|}{$2$}  \\ \hline

 cache controller &  $381$  &   $67$       & \multicolumn{1}{c|}{$91.21\%$}   & \multicolumn{1}{c|}{$100.00\%$}       & \multicolumn{1}{c|}{$91.21\%$}  & \multicolumn{1}{c|}{$1 $}   & \multicolumn{1}{c|}{$1 $}    & 
 \multicolumn{1}{c|}{$1$}   & 
 \multicolumn{1}{c|}{$2$}  & \multicolumn{1}{c|}{$1$}   & \multicolumn{1}{c|}{$2$}    &  \multicolumn{1}{c|}{$1 $}    &   \multicolumn{1}{c|}{$1$}   &    \multicolumn{1}{c|}{$1$}  &  \multicolumn{1}{c|}{$2 $} \\ \hline

\end{tabular}
\label{tab:results_main}
}
\end{table*}
We consider a set of 6 modules from the BRISC-V RTL code base. 
A procedural block of an RTL code specifies the behavioral property of the code. It comprises statements like assignment statements, conditional statements, and state transition statements. In all of the experiments, we consider always blocks, and in most cases of the chosen modules, the lines of code in each block are around 200-300 lines. This is typically 
representative of a situation where a designer can localize a given 
bug to a small region of code or an always block, but is unable to 
pinpoint it. This is also an advantage of the proposed methodology, as the prior work assumed that the line with the bug is known a priori~\cite{ACMTODES-khushboo}. Given that we have the assertions for the RTL, we could narrow down the region of the bug to 200-300 lines of code for all the introduced mutations. 
The details of the benchmark modules used for the evaluation are furnished 
in the following discussion.
\begin{itemize}[leftmargin=*]
    \item  \textbf{\textit{L1\_bus\_controller}}: Comprises $14$ states to make the interface between snooper, L1 cache memory and bus for communication.

    \item \textbf{\textit{coherence\_controller}}: The purpose of the coherence controller in BRISC-V is to monitor the bus access between the caches and the shared memory. Only one cache at a time gets bus access to communicate with the shared memory. There are 7 states to capture various scenarios that can occur during execution. 

    \item \textbf{\textit{snooper}}: Snooper observes the operations performed by the other cores through its 7 states, when to invalidate its own data based on the write operation by other cores.  

    \item \textbf{\textit{Lxcache\_controller}}: Takes care of the L2 cache through its $9$ states, also manages valid, invalid, inclusive bits to exercise the coherence protocol (MESI in this case).

    \item \textbf{\textit{Lxbus\_interface}}: Bridges between bus and L2 cache with its $11$ states to coordindate the flow of message, address and data.

    \item \textbf{\textit{cache\_controller}}: The L1 cache of each core has a cache controller module that controls the required address, data, and message flow between the communicating modules. There are 15 states in the cache controller module to preserve the correctness of the whole system to implement the MESI protocol.
\end{itemize}



\subsection{Bug Insertion}
To evaluate the effectiveness of our methodology, we leverage mutation testing, where we create mutations of ${\mathcal R}$. These mutations act as bugs in our design. 
To ensure that the mutations in code lines 
are inserted at random locations, we use the LLM itself to create them. 
It has to be noted that the mutations are created in a separate session 
so that the LLM does not remember its history during the repair step. The mutations inserted in the correct code are syntax preserving, while semantically non-preserving~\cite{accurate}, which results in altering the functionality of the code 
that was captured in the correct assertions. As mentioned earlier, a single 
mutation is inserted in a single line of the RTL code, resulting in 
a semantic change that causes one or more of ${\mathcal P}$ to be violated.

The mutations engage both the sequential and combinational elements of 
the code. Specifically, we work on the same set of mutations defined 
in~\cite{ACMTODES-khushboo}, as summarized in Table~\ref{tab:mutable}, with 
one additional type (last item), as discussed below. Prompt~3 is used to 
generate the mutations.  

The list of mutations shown in Table~\ref{tab:mutable} describes the set of $10$ mutations that we have carried out for our experiments. Column 1 states the serial number, the second column mentions the name of the mutation, and in the third column, a short description about the mutation is provided.


\textbf{Nested if-else Mutation:} In this mutation~\cite{vert}, the else statement corresponding to an if statement is misaligned. 
Evidently, this leads to a 
semantic mismatch with respect to the correct functionality. 
As shown in Fig.~\ref{fig:figNestedIf}, the correct code shows the intended nested if-else alignment, where the \texttt{"if (dirty)"} condition is nested in the \texttt{"if (hit)"} block, but we see in the mutation column that the \texttt{"if (dirty)"} condition is placed with the \texttt{"else"} statement, and the nested if-else construct is broken. This may lead to 
assertion violations, which are important for our repair task. For each mutation type, we insert an instance in the given RTL ${\mathcal R}$ using the prompt shown in Prompt~3 and insert the bugs in the design.

\begin{table}[tb]
\centering

\caption{Two mutations introduced in each BRISC-V module (first of type A and second of type B)}
\label{tab:results_2M}
\resizebox{0.7\linewidth}{!}{%
 

\begin{tabular}{|c|c|c|}
\hline
 \begin{tabular}[c]{@{}c@{}}Mutation Types \\ A : B \end{tabular}    & \begin{tabular}[c]{@{}c@{}}\# bugs repaired \\ at \textit{Stage 1}\end{tabular} & \begin{tabular}[c]{@{}c@{}}\#bugs repaired\\  at \textit{Stage 2}\end{tabular} \\ \hline
Conditional : Connection & 2                                                                      & 0                                                                     \\ \hline
Arithmetic : Conditional & 2                                                                      & 0                                                                     \\ \hline
Arithmetic : Connection  & 1                                                                      & 1                                                                     \\ \hline
Logical : Arithmetic     & 1                                                                      & 1                                                                     \\ \hline
Logical : Conditional    & 0                                                                      & 2                                                                     \\ \hline
Logical : Connection     & 0                                                                      & 2                                                                     \\ \hline
Logical : Relational     & 1                                                                      & 1                                                                     \\ \hline
Relational : Arithmetic  & 0                                                                      & 2                                                                     \\ \hline
Relational : Conditional & 2                                                                      & 0                                                                     \\ \hline
Relational : Connection  & 0                                                                      & 2                                                                     \\ \hline
\end{tabular}
}
\end{table}

\begin{table}[tb]
\centering
\caption{Three mutations introduced in each BRISC-V module (first of type A, second of type B, third of type C)}
\resizebox{0.8\linewidth}{!}{%
 
\begin{tabular}{|c|c|c|}
\hline
 \begin{tabular}[c]{@{}c@{}}Mutation Types \\ A : B : C \end{tabular}    & \begin{tabular}[c]{@{}c@{}}\# bugs repaired \\ at \textit{Stage 1}\end{tabular} & \begin{tabular}[c]{@{}c@{}}\#bugs repaired \\ at \textit{Stage 2}\end{tabular} \\ \hline
Logical :  Conditional :  Connection  & 0                                                                      & 3                                                                     \\ \hline
Logical : Relational :  Conditional   & 3                                                                      & 0                                                                     \\ \hline
Relational : Conditional : Connection & 2                                                                      & 1                                                                     \\ \hline
Logical : Relational : Connection     & 2                                                                      & 1                                                                     \\ \hline
Relational : Connection : If \_Else   & 2                                                                      & 1                                                                     \\ \hline
\end{tabular}
\label{tab:results_3M}
}

\end{table}

\begin{table}[tb]
\centering
\caption{Four mutations introduced in each BRISC-V module (first of type A, second of type B, third of type C, and fourth of type D)}
\label{tab:results_4M}
\resizebox{0.9\linewidth}{!}{%
\begin{tabular}{|c|c|c|}
\hline
 \begin{tabular}[c]{@{}c@{}}Mutation Types \\ A : B : C : D \end{tabular}   & \begin{tabular}[c]{@{}c@{}}\# bugs repaired \\ at \textit{Stage 1}\end{tabular} & \begin{tabular}[c]{@{}c@{}}\#bugs repaired\\  at \textit{Stage 2}\end{tabular} \\ \hline
Logical : Relational : Conditional : Arithmetic & 2                                                                      & 2                                                                     \\ \hline
Logical : Relational : Conditional : Connection & 3                                                                      & 1                                                                     \\ \hline
Logical : Relational : Connection : Arithmetic  & 3                                                                      & 1                                                                     \\ \hline
\end{tabular}
}
\end{table}



\begin{table}[tb]
\centering
\caption{8 different types of  mutations in L1 bus Interface and Snooper module, all at once, 1: Repaired in \textit{Stage 1}, 2: Repaired in \textit{Stage 2}}
\label{tab:results_multiM_L1bus_snooper}
\resizebox{0.65\linewidth}{!}{%
\begin{tabular}{|c|cc|}
\hline
                 & \multicolumn{2}{c|}{Stage at which the bug was repaired} \\ \hline
Mutation Types   & \multicolumn{1}{c|}{L1 bus interface}      & Snooper     \\ \hline
Relational       & \multicolumn{1}{c|}{1}            & 1                    \\ \hline
Arithmetic       & \multicolumn{1}{c|}{1}            & 1                    \\ \hline
Conditional      & \multicolumn{1}{c|}{1}            & x                    \\ \hline
Stuck\_At\_Fault & \multicolumn{1}{c|}{1}            & 2                    \\ \hline
Connectivity     & \multicolumn{1}{c|}{1}            & 2                    \\ \hline
Constant         & \multicolumn{1}{c|}{2}            & 2                    \\ \hline
Slicing          & \multicolumn{1}{c|}{2}            & 1                    \\ \hline
Nested If\_else  & \multicolumn{1}{c|}{2}            & 2                    \\ \hline
\end{tabular}
}
\end{table}
\subsection{Results and Findings}\label{sec:resultsandfind}
\noindent
We implement an immediate prior art~\cite{ACMTODES-khushboo} (referred to as 
\textit{Stage 1} in our experiments) on LLM-aided 
RTL repair, and provide comparative results. As seen from our experiments, the presence of assertions indeed helps the fixing task.
In most cases, after \textit{Stage 1}, the LLM response is to suggest possible errors that can be present in the code. For a few cases, it is able to firmly say that this is the buggy line with a justification. Consider the case of a constant mutation inserted in the Reset block of the code, where \texttt{REQ1\_read} is assigned \texttt{1'b1}, with all other signals assigned to \texttt{1'b0} in the reset block. In this case, the LLM is able to find the bug due to prior knowledge of the reset cycle. 
In some cases, as well, it is able to identify the bug, but is not able to patch it in \textit{Stage 1}. An example session is shown in Listing~\ref{code1} with one mutated line where the index of \texttt{curr\_offset} got mutated. Listing~\ref{code2} shows the patch suggested by LLM after \textit{Stage 1}, which identifies that the bug is in the offset value, but the patch suggested produces failing assertions. At \textit{Stage 2}, when a list of failing assertions is given, the correct offset is produced (shown in Listing~\ref{code3}),  resulting in all passing assertions. This clearly demonstrates the advantage of having assertions. 

\begin{lstlisting} [frame=single,caption={Sample code with mutated line},label=code1]
r_cache_data_out[j] <= (j>=curr_offset & 
j < (curr_offset+req_words))?
curr_data[j] : r_cache_data_out[j];\end{lstlisting}

\begin{lstlisting} [frame=single,caption={Patch suggested by LLM after \textit{Stage 1}},label=code2,
belowskip=2pt]
FIX: use logical && and index curr_data with word_counter offset
r_cache_data_out[j] <= ((j >= curr_offset)
&& (j < (curr_offset + req_words))) ?
curr_data[word_counter + j]
:r_cache_data_out[j];\end{lstlisting}

\begin{lstlisting} [frame=single,caption={Patch suggested by LLM after \textit{Stage 2}},label=code3,aboveskip=2pt,
belowskip=2pt]
r_cache_data_out[j] <= (j>=curr_offset &
j<(curr_offset+req_words))? 
curr_data[j-curr_offset]
:r_cache_data_out[j];\end{lstlisting}

\noindent
We now discuss the overall results obtained using our methodology. Table~\ref{tab:results_main} shows the following descriptions. The first column shows the name of the module from the BRISC-V module. The second column and the third column give the number of lines of code and the number of assertions, respectively. The fourth column gives the three coverage metrics. \emph{Formal Coverage} (FC): combines \emph{Stimuli Coverage} (SC) and \emph{Checker Coverage} (CC). An item is formally covered if it is covered by SC and CC. SC shows the reachability of the formal engines. CC shows the \emph{Cone of Influence} (COI) coverage of the generated assertions. In general, the higher the coverage, the better the quality of the assertions. The last column describes the mutation types in each sub-column. A triplet in each sub-column indicates the outcome after the individual stages: \textit{Stage 1}, \textit{Stage 2}. 

In total, 60 cases (6 modules, and 10 types of mutations) result from Table~\ref{tab:mutable}. However, we could not introduce the slicing bug in the Coherence Controller IP, so we have $59$ cases. This acts as a baseline for our work, which only uses the LLM for bug fixing. 
From Table~\ref{tab:results_main}, we see that $28$ bugs are correctly patched after \textit{Stage 1}. This can be considered as a \textit{baseline}, as the prior works do not have assertions available for the selected IP. In addition, when we tried generating assertions using the existing methodology, the quality of the assertions were poor as discussed in Section~\ref{sec:assertion}.  
After \textit{Stage~2}, we were able to patch $29$ bugs. Overall, after all the stages, $1$ of the bug was not patched. This gives us an overall success rate of $98\%$. For modules like \emph{Cache Controller} with $(381)$ lines and with $68$ assertions, the number of tokens ranges from $4$k  to $4.5$k based on the number of stages required to patch the bug. The JasperGold runtime ranges from $135ms$ in the \emph{cache coherence} module to $1273ms$ in the \emph{L1 Bus Interface} module. This also shows the effectiveness of a closed-loop LLM-assisted methodology, which is able to provide RTL bug repairs with formal guarantees and a significantly high bug repair rate.

After achieving a success rate of $98\%$, we applied our methodology to fix multiple bugs across multiple lines. In the first step, we consider a smaller module (number of lines is $171$) like \emph{Coherence Controller}, and introduce a smaller number of mutations there. Firstly,  5 types
of mutations (Logical, Conditional, Connection, Arithmetic, and Relational) are considered, and  $\binom{5}{2} = 10 $  different mutation combinations are introduced on one module for our experiment, the results are reported in Table ~\ref{tab:results_2M}. Table ~\ref{tab:results_2M} contains 3 columns. Column 1 shows the types of mutation introduced, and the other two columns show the number of bugs repaired in each stage. Column 2 shows the number of bugs fixed in \textit{Stage 1}, and  Column 3 shows the number of bugs fixed in \textit{Stage 2}.
Similarly, Tables~\ref{tab:results_3M} and ~\ref{tab:results_4M}, show the results after introducing 3 and
4 different types of mutations in the module at a time. We also see that, within the \textit{Stage 2}, all bugs are repaired. In each of the experiments reported in Tables~\ref{tab:results_2M},~\ref{tab:results_3M}, and ~\ref{tab:results_4M}, we get $100\%$ success in bug repairing. Thus, the proposed methodology is feasible for the correction of multi-line bugs of different types simultaneously.


Lastly, we considered relatively large modules (number of lines is more than 300), like \emph{Snooper} and \emph{L1 bus interface}, and introduced $8$ different types of mutations all at the same time as shown in Table~\ref{tab:results_multiM_L1bus_snooper}. Column 1 shows the $8$ types of different mutations, Column 2 and Column 3 show the \textit{Stage} number at which that bug gets repaired for the \emph{L1 bus interface} and \emph{Snooper}, respectively. We see that in Table~\ref{tab:results_multiM_L1bus_snooper}, except for one case, we were able to correct all the mutations in the two \textit{Stages}, giving us a $93\%$ success rate here. This clearly highlights that the proposed method can handle multiple bugs from several classes simultaneously and correct them.

Overall, we showed that the proposed methodology, in which the model checker aids the LLM, can be used for the bug patching of larger designs with multi-line bugs.

\section{Conclusion}\label{sec:conc}
In this work, we proposed the first attempt towards LLM-assisted automated RTL bug repair of single-line and multi-line bugs with assertions and a model checker in the loop. 
The methodology consists of two distinct steps and follows a closed-loop principle to build a context for the LLM from the assertions and the output of the model checker.
Consequently, it provides a bug fix along with trustworthy grounding and formal guarantees.
We used mutation testing to perform extensive experiments on BRISC-V.
Our proposed method was able to fix up to $98\%$ of single-line RTL bugs and 93\% of the multi-line bugs. The fixes were formally verified using the model checker.  In addition, we believe that our methodology can be further refined to identify the few bugs that were not corrected. 
\makeatletter
\newcommand*\mysize{%
  \@setfontsize\mysize{7.5}{8.5}%
}
\makeatother
\IEEEtriggeratref{1}
\IEEEtriggercmd{\mysize}
\renewcommand\refname{References\vspace{-0.1cm}}
\bibliographystyle{IEEEtran}
\bibliography{references}

\end{document}